\documentstyle[epsfig]{mn}

\newif\ifAMStwofonts
\AMStwofontstrue                % <- comment this out if AMS fonts not available
\DeclareMathVersion{bold}                    % <- seems to be needed with mathptm
\DeclareMathAlphabet{\mathbfit}{OT1}{cmr}{bx}{it}
\SetMathAlphabet\mathbfit{bold}{OT1}{cmr}{bx}{it}
\DeclareMathAlphabet{\mathbfss}{OT1}{cmss}{bx}{n}
\SetMathAlphabet\mathbfss{bold}{OT1}{cmss}{bx}{n}
\ifAMStwofonts
  \ifCUPmtlplainloaded \else
    \DeclareSymbolFont{UPM}{U}{eur}{m}{n}
    \SetSymbolFont{UPM}{bold}{U}{eur}{b}{n}
    \DeclareSymbolFont{AMSa}{U}{msa}{m}{n}
    \DeclareMathSymbol{\upi}{0}{UPM}{"19}
    \DeclareMathSymbol{\umu}{0}{UPM}{"16}
    \DeclareMathSymbol{\upartial}{0}{UPM}{"40}
    \DeclareMathSymbol{\leqslant}{3}{AMSa}{"36}
    \DeclareMathSymbol{\geqslant}{3}{AMSa}{"3E}

    \let\leq=\leqslant 
     \let\ge=\geqslant
  \fi
\fi
\title[Wavelets and non-Gaussianity]
{The discriminating power of wavelets to detect non-Gaussianity in the
CMB}
\author[Barreiro \& Hobson]{R.B.~Barreiro and M.P.~Hobson\\
Astrophysics Group, Cavendish Laboratory, Madingley Road,
Cambridge CB3 0HE, UK }
\date{Accepted ???. Received ???; in original form ???}

\begin{document}
\maketitle
\begin{abstract}
We investigate the power of wavelet techniques in detecting
non-Gaussianity in the cosmic microwave background (CMB).
We use the method
to discriminate between an inflationary and a cosmic strings model
using small simulated patches of the sky. We show the importance of the
choice of a good test statistic in order to optimise the
discriminating power of the wavelet technique. In particular,
we construct the Fisher discriminant function, which
combines all the information available in the different wavelet
scales. We also compare the
performance of different decomposition schemes and wavelet bases.   
For our case, we find that the Mallat and {\it \`a trous} algorithms
are superior to the 2D-tensor wavelets. Using this technique,
the inflationary and strings models are
clearly distinguished even in the presence of a superposed 
Gaussian component with twice the rms amplitude of the original cosmic
string map.
\end{abstract}

\begin{keywords}
methods: data analysis-cosmic microwave background.
\end{keywords}

\section{Introduction}

The Cosmic Microwave Background (CMB) is currently one of the most
powerful tools of cosmology to investigate the theories of structure
formation in the Universe. In particular, the standard inflationary
model predicts Gaussian fluctuations whereas alternative theories such
as topological defects scenarios give rise to non-Gaussianity in the
temperature field. Therefore, the study of the Gaussian character of
the CMB will allow us to discriminate between these two competitive
theories of structure formation. In addition, future high-resolution
maps should be investigated for any trace of non-Gaussianity that may be
imprinted by foregrounds or systematics.

In order to test the Gaussianity of the CMB, a large number of methods
have been proposed (for a review see Barreiro
2000). Perhaps the most straightforward of these is the measurement of
the skewness and kurtosis of the distribution of the pixel
temperatures in the map (Scaramella \& Vittorio 1991). More
complex methods include properties of hot and cold spots
(Coles \& Barrow 1987; Mart\'\i nez-Gonz\'alez et al. 2000; Barreiro,
Mart\'\i nez-Gonz\'alez \& Sanz 2001), extrema correlation function
(Naselsky \& Novikov 1995,
Kogut et al. 1996, Barreiro et al. 1998, Heavens \& Sheth 1999),
Minkowski functionals (Coles 1988, Gott et al. 1990, Kogut et al. 
1996), bispectrum analysis (Ferreira et al. 1998, Heavens 1998,
Magueijo 2000), multifractals (Pompilio et al. 1995) and partition function 
(Diego et al. 1999, Mart\'\i nez-Gonz\'alez et al. 2000).

In addition, tests of Gaussianity based on wavelets have become quite
popular in the last years. Ferreira, Magueijo \& Silk (1997) propose a
set of statistics using cumulants and defined
in wavelet space.
Hobson, Jones \& Lasenby (1998) investigate the power of the cumulants
of the distribution of wavelet coefficients at each scale on detecting
the Kaiser-Stebbins effect using CMB simulations of small patches of
the sky. A similar method was introduced by Forni \& Aghanim (1999)
and tested using simulated CMB maps containing secondary anisotropies 
(Aghanim \&
Forni 1999) and on the COBE data (Aghanim, Forni \& Bouchet 2001).
Pando, Valls-Gabaud \& Fang (1998) calculated the skewness, kurtosis
and scale-scale correlation coefficients of the COBE data in the
QuadCube pixelisation using planar orthogonal wavelets. This study was
extended by Mukherjee et al (2000). Barreiro et al (2000) applied the
same method to the COBE data in HEALPix pixelisation using spherical
wavelets. 

This paper is an extension of the work done by Hobson et al (1998)
(HJL, hereinafter). As in HJL, the fourth
cumulant of the distribution of wavelet coefficients is calculated at
each wavelet scale using high-resolution CMB simulations of small
patches of the sky for Gaussian and non-Gaussian simulations. 
However, in this work, all this information is combined into the Fisher
discriminant function in order to get a more statistically significant
result. In addition, we compare the performance of three different
ways of constructing 2-dimensional wavelets: tensor
(as in HJL), Mallat and the {\it \`a trous} algorithm.
Moreover, some of the studied wavelets are 
invariant under rotations of 90, 180 and 270 degrees of the original
data, which solves the earlier problem of orientation sensitivity.

%In this paper, we extend the work done by Hobson et al (1998) (HJL,
%hereinafter). The main extension is the comparison of the performance
%of different types of wavelets and the use of the Fisher discriminant
%function, a test statistic that
%combines all the information contained at the different wavelet
%scales. In addition, the orientation sensitivity problem is also
%addressed by using wavelets that are invariant under rotations of 90,
%180 and 270 degrees of the original data.

The outline of the paper is as follows. \S\ref{2dwavelets} gives a
brief introduction to the three different constructions of 2-dimensional
wavelets. In \S\ref{test} our non-Gaussianity test is explained in
detail, also pointing out the interest of the Fisher discriminant
function. The results of applying this test to simulated data
are given in \S\ref{results}. Finally, our conclusions are presented
in \S\ref{concl}.

\section{The discrete wavelet transform}
\label{2dwavelets}
The wavelet transform has been extensively discussed
elsewhere (e.g. Daubechies 1992). 
In particular, a clear introduction to wavelets is given by Burrus,
Gopinath \& Guo (1998). We therefore give only a brief
introduction, pointing out the differences between the three
considered decomposition schemes.
The basis of the wavelet transform is to decompose the considered
signal (the temperature field in our case) into a set of 
wavelet coefficients. Each of them contain information about the signal at a
given position and scale. Therefore, wavelets
allow one to study the structure of an image at
different scales and at the same time keep localisation.
There is not a unique choice for a wavelet basis. Moreover,
different algorithms schemes can be used to construct 2-dimensional
wavelets. In particular, we consider the tensor, 
Mallat (or multiresolution) 
and {\`a trous} algorithms.

\subsection{Tensor algorithm}
\label{tensor}
This is the type of wavelets used in the analysis of HJL, where a more
detailed description can be found. In 1-dimension, the wavelet basis
is constructed from dilations and translations of the mother (or
analysing) wavelet function $\psi$ and a second related function
called the father (or scaling) wavelet function:
\begin{eqnarray}
\psi_{j,l} & = & 2^{\frac{j-J}{2}}\psi\left( 2^{j-J}x-l\right) \, , 
\nonumber \\
\phi_{j,l} & = & 2^{\frac{j-J}{2}}\phi\left( 2^{j-J}x-l\right) \, , 
\end{eqnarray}
where $ 0 \ge j \ge J-1$ and $ 0 \ge l \ge 2^J-1 $ are integer denoting 
the dilation and translation
indices, respectively, and $2^J$ is the number of pixels of 
the considered discrete signal $f(x)$.

In order to construct a real, orthogonal and compactly-supported wavelet
basis, such as the Haar and
Daubechies 4 wavelets used in this work, these functions must satisfy
some highly-restrictive mathematical relations (Daubechies 1992). 
The discrete signal $f(x)$ can thus be written as
\begin{equation}
f(x_i)= a_{0,0}\phi_{0,0}(x_i) + \sum_j \sum_l d_{j,l}\psi_{j,l}(x_i)
\, ,
\end{equation}
being $a$, $d$ the approximation and detail wavelet coefficients
respectively. These coefficients can be found in a recursive way
starting from the data vector $f(x_l) \equiv a_{J,l}, l=0,..,2^{J}-1$
\begin{eqnarray}
a_{j-1,l} & = & \sum_m h(m-2l)a_{j,m} \nonumber \\
d_{j-1,l} & = & \sum_m g(m-2l)a_{j,m} \, .
\end{eqnarray}
where $h$, $g$ are the low and high-pass filters associated to the scaling
and wavelet functions through the refinement equation (see e.g. Burrus
et al. 1998). At each iteration, the vector of
length $2^j$ is split into $2^{j-1}$ detail components and $2^{j-1}$
smoothed components. The decimated smoothed components are then used
as input for the next iteration to construct the detail coefficients
at the next larger scale. This produces a number of wavelet
coefficients equal to the original number of pixels.
As the index $j$ increases from 0 to $J-1$, the wavelet
coefficients correspond to the structure of the function on
increasingly smaller scales, with each scale a factor of 2 finer than
the previous one. 

The extension of the DWT to 2-dimensional signals is obtained 
by taking tensor products of the one-dimension wavelet basis as
follows 
\begin{eqnarray*}
\phi_{0,0;l_1,l_2}(x,y) & = & \phi_{0,l_1}(x)\phi_{0,l_2}(y) \\
\zeta_{j_1,0;l_1,l_2}(x,y) & = & \psi_{j_1,l_1}(x)\phi_{0,l_2}(y) \\
\xi_{0,j_2;l_1,l_2}(x,y) & = & \phi_{0,l_1}(x)\psi_{j_2,l_2}(y) \\
\psi_{j_1,j_2;l_1,l_2}(x,y) & = & \psi_{j_1,l_1}(x)\psi_{j_2,l_2}(y).
\end{eqnarray*}
If the two-dimensional pixelised image has dimensions $2^{J}\times
2^{J}$ then we have 
$0 \leq j_1 \leq J-1$ and $0 \leq l_1 \leq 2^{j_1}-1$, and similarly
for $j_2$ and $l_2$.

With this wavelet decomposition scheme, the wavelet transform is
performed independently in both directions. By analogy with the
one-dimensional case, as $j_1$ ($j_2$) increases the wavelet coefficients
contain information about the horizontal (vertical) structure in the image on
increasingly smaller scales. 
In particular, the scale corresponding to each region is
approximately $\sim$pixel~size$\times 2^{J-j_i}$ in each direction. 
Therefore, regions with $j_1=j_2$ contain
coefficients of two-dimensional wavelets that represent the image at
the same scale in the horizontal and vertical directions. Conversely, 
domains with $j_1 \ne j_2$ contain wavelet coefficients describing the 
image on different scales in the two directions. 

\subsection{Mallat algorithm}
\label{mallat}
For the previous wavelets, the wavelet transform is applied separately
for each direction of the image. This means that some of the domains
contain mixing of structure with different scales $j_1$ and $j_2$ in
the horizontal and vertical directions. However, 
it is possible to use one-dimensional bases to construct
two-dimensional bases that do not mix scales and can be described in
terms of a single scale index $j$ (Mallat 1989). 
At scale $j$, these bases are given by 
\begin{eqnarray*}
\phi_{j;l_1,l_2}(x,y) & = & \phi_{j,l_1}(x)\phi_{j,l_2}(y) \\
\psi^{\rm H}_{j;l_1,l_2}(x,y) & = & \psi_{j,l_1}(x)\phi_{j,l_2}(y) \\
\psi^{\rm V}_{j;l_1,l_2}(x,y) & = & \phi_{j,l_1}(x)\psi_{j,l_2}(y) \\
\psi^{\rm D}_{j;l_1,l_2}(x,y) & = & \psi_{j,l_1}(x)\psi_{j,l_2}(y).
\end{eqnarray*}
In this wavelet decomposition, the temperature map (scale $J$) 
is decomposed into
an approximation image (that is basically a lower resolution version 
of the original image) plus three details images. These detail
coefficients contain horizontal, vertical and diagonal structure at
scale $J-1$ that encode the differences between the original and lower
resolution images. The same
decomposition is applied then to the approximation image, generating
a lower resolution image and another three details at scale
$J-2$. This process is repeated down to the lowest resolution level
considered. The scale of the structure contained by a wavelet
coefficient with index $j$ correspond approximately to 
pixel~size$\times 2^{J-j}$
As for the tensor case, the wavelet basis is non-redundant and
orthogonal. Thus, the
number of wavelet coefficients is the same as the number of pixels.
We also note that the diagonal wavelet coefficients obtained with the 
Mallat algorithm
are the same as those with $j_1=j_2$ in the tensor decomposition.

Although different orthogonal wavelet bases were tested for the tensor
and Mallat algorithms, we present our results for Haar and Daubechies 4 
(see \S\ref{results}). We note that it is not possible to
construct a real orthogonal wavelet basis with compact support whose
wavelet basis functions are also symmetric. Therefore the
wavelet transform will not be invariant under rotations. However, the
Haar wavelet functions are antisymmetric. This means that our test will be
invariant under rotations of 90, 180 and 270 degrees of the data with respect
to the original orientation when using the Haar wavelets basis.

\subsection{{\it \`A trous} algorithm}
\label{atrous}
The {\it \`a trous} (`with holes') algorithm (Holschneider \&
Tchamitchian 1990; Shensa 1992) 
is a redundant non-orthogonal wavelet transform. 
Given a data vector $f(x_l) \equiv c_{J,l}, l=0,...,2^J-1$, the approximation 
coefficients for the next lower resolution are obtained as
\begin{equation}
a_{j-1,l}=\sum_m h(m)a_{j,l+2^{(J-j)}m}
\end{equation}
i.e., each step is a convolution of the data with the filter $h$ and
varying step sizes $2^{J-j}$. 
The function $h$ is a discrete low pass filter associated to a scaling function
$\phi(x)$ that can be derived from a refinement relation (see
e.g. Starck \& Murtagh 1994).
The corresponding detail coefficients are simply obtained by
subtracting the smoother image from the original one:
\begin{equation}
d_{j-1,i}=a_{j,k}-a_{j-1,k},
\end{equation}
Since no decimation is carried out between consecutive filter steps,
the number of wavelet coefficients at each scale is $2^J$. Note that
the reconstruction of the signal is particularly simple in this case,
obtained by adding the detail wavelet coefficients at all scales plus
the approximation coefficients at the lowest considered resolution:
\begin{equation}
c_{J,l}=c_{0,l}+\sum_{j=0}^{J-1}w_{j,l}
\end{equation}

We have chosen a $B_3$-spline for the scaling function, which leads to
a convolution with a mask
$(\frac{1}{16},\frac{1}{4},\frac{3}{8},\frac{1}{4},\frac{1}{16})$ 
in 1-dimension and

\begin{equation}
\left(
\begin{array}{ccccc}
\frac{1}{256} & \frac{1}{64} & \frac{3}{128} & \frac{1}{64} &
\frac{1}{256} \\
\frac{1}{64} & \frac{1}{16} & \frac{3}{32} & \frac{1}{16} &
\frac{1}{64} \\
\frac{3}{128} & \frac{3}{32} & \frac{9}{64} & \frac{3}{32} &
\frac{3}{128} \\
\frac{1}{64} & \frac{1}{16} & \frac{3}{32} & \frac{1}{16} &
\frac{1}{64} \\
\frac{1}{256} & \frac{1}{64} & \frac{3}{128} & \frac{1}{64} &
\frac{1}{256} \\
\end{array}
\right)
\end{equation}
in 2-dimensions (e.g. Starck \& Murtagh 1994).
Note that, due to the symmetry of the mask, the {\`a trous} wavelet
transform is invariant under rotations of the data of 90, 180 and
270 degrees with respect to the original orientation.

\section{The non-Gaussianity test}
\label{test}
The non-Gaussianity test used in this paper is basically the same as that
of HJL. The main idea is to perform a wavelet
transform of the (possibly) non-Gaussian image and obtain
certain statistic of the wavelet coefficients in each 
region of the transform. In
particular, we estimate the fourth cumulant $\hat\kappa_4$ for each
scale. This quantity is calculated using $k-$statistics as explained in
HJL. We repeat this test for a large number (10000) of non-Gaussian and
Gaussian realisations and then compare the distribution of the fourth
cumulant spectra $\hat\kappa_4(r)$ obtained from the two different
populations. This is different to HJL where only a single non-Gaussian
map was used to test the power of the wavelet technique (see \S\ref{results}).
Each Gaussian map is obtained from one of the non-Gaussian maps by
randomising its phases. Therefore, Gaussian and non-Gaussian maps
share the same power spectrum by construction. 
This is important to ensure that any
possible differences found between both populations are due to the
non-Gaussian character of the test map rather than differences in the
power spectrum.

As discussed in HJL, due to edge effects, 
not all wavelet coefficients can be used in the analysis.
We need to discard those coefficients obtained from
wavelet functions that cross the boundary of the image to avoid
misidentifying these borders as non-Gaussian features. To identify
the affected wavelet coefficients,
we perform a wavelet transform of a map consisting of zeroes
except for a border of non-zero pixels on the map boundary. The
resulting non-null wavelet coefficients correspond to edge-crossing
basis functions and are therefore ignored in the subsequent analysis.

One can, of course, repeat the test using cumulants other than
$\kappa_4$. We have obtained results for $\kappa_3$, but find it to be
a less discriminating statistic than in the case studied here.

\subsection{The discriminating power of a test}

In HJL, the significance of the test was given by the level of the
highest detection of non-Gaussianity.
However, a single detection of non-Gaussianity may not be statistically
significant if all the other estimators are consistent with a Gaussian
distribution. Conversely, when the level of confusion is high, it may
happen that no single statistic detects non-Gaussianity at a high
significance level. Nevertheless, if we use all the available statistics
together, we may find a systematic
displacement with respect to the Gaussian distribution, which would
allow us to discriminate between both models.
Therefore, a more statistically rigorous
analysis should take into account all the information provided by
the different regions of the wavelet transform.
In order to do this, we need to construct a test statistic that
maximises the discriminating power between the two models.

Suppose that given a set of $n$ measurements ${\mathbf x}=(x_1,...,x_n)$,
we want to check the validity of a hypothesis
$H_0$, usually called the null hypothesis, 
against an alternative hypothesis $H_1$. Each hypothesis
specifies a joint probability $f({\mathbf x}|H_0)$ and 
$f({\mathbf x}|H_1)$ for obtaining the particular values of our data. 
In our case, the measurements correspond to the
value of the fourth cumulant at each wavelet scale and we want to
test if our sample is drawn from a Gaussian model
or from a cosmic strings scenario. In order to compare the level of
agreement between the measured data and the considered hypotheses, it
is useful to construct a function of the measured variables called a
test statistic $t({\mathbf x)}$. In principle, this statistic can be
a multidimensional vector, but we consider only the case when $t$ is a
scalar, since this greatly simplifies the problem by reducing the
number of dimensions of the data whilst keeping enough information to
discriminate between the considered hypotheses. 
The probability of obtaining a certain value of $t$ will be given by
$g(t|H_0)$ if $H_0$ is true and by
$g(t|H_1)$ if the correct hypothesis is $H_1$.
\begin{figure}
{\epsfig{file=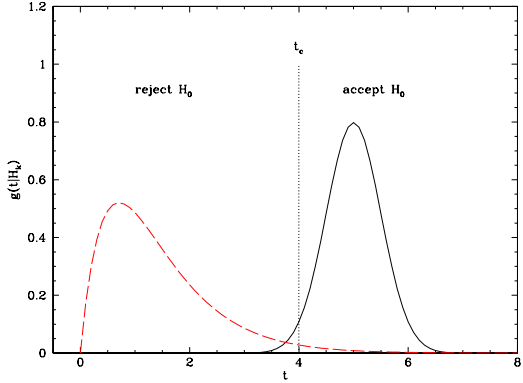,width=5.5cm,angle=-90}}
\caption{An example of the discriminating power of a test between two 
hypotheses $H_0$ and $H_1$. The solid line corresponds to the
probability $g(t|H_0)$ of obtaining a certain value of $t$ if $H_0$ is
true, whereas the dashed line gives the same probability $g(t|H_1)$ 
if $H_1$ is the correct hypothesis.}
\label{hipotesis}
\end{figure}
In order to reject or accept the hypothesis $H_0$, one can define a
critical value for $t$ (see Fig.~\ref{hipotesis}). 
The critical value $t_c$ is defined so that the probability of
observing $t$ under the assumption of the hypothesis $H_0$ is less than
a certain value $\alpha$, which is called the significance level of
the test. In the example of Fig.~\ref{hipotesis} this would be given by 
\begin{equation}
\alpha=\int_{-\infty}^{t_{\rm c}}g(t|H_0)dt ~.
\end{equation}
Thus, there is a probability $\alpha$ of rejecting $H_0$ when it is
actually true. This
is called an error of the first kind. An error of the second kind
occurs when $t$ falls in the acceptance region but it was actually
drawn from an alternative hypothesis $H_1$. In our example, in
Fig.~\ref{hipotesis}, the probability of such an error happening is
given by
\begin{equation}
\beta=\int_{t_{\rm c}}^{\infty}g(t|H_1)dt ~.
\end{equation}
The quantity $p \equiv 1-\beta$ is called
the power of the test to discriminate
against the alternative hypothesis $H_1$ at the considered
significance level.
In our wavelet analysis, we will compare the values of $p$ 
for a significance level of $\alpha=0.01$ for each of
the studied cases.

\subsection{The $\chi^2$ test}
\label{chitest}
The key issue now is the choice of a test statistic. One possibility
is to uncorrelate the variables and then obtain a 
standard $\chi^2$ statistic. The
procedure is as follows. First, the empirical covariance matrix $V$ of
the variables 
is obtained from a large number of Gaussian simulations. Then, a new
set of variables ${\mathbf x}$ is constructed by a rotation:
\begin{equation}
{\mathbf x^\prime}=S^T{\mathbf x}
\end{equation}
where $S$ is a matrix whose columns are the normalised 
eigenvectors of the covariance matrix $V$. The same 
rotation is then applied to the non-Gaussian variables. Finally, the
$\chi^2$ statistic is obtained for each Gaussian and non-Gaussian
realisation by calculating
\begin{equation}
\chi^2=\sum_i\left(\frac{x^\prime_i-\mu^\prime_i}{\sigma^\prime_i}\right)^2
\label{eqchi2}
\end{equation}
where $x^\prime_i$ corresponds to the $i$th (Gaussian or
non-Gaussian) rotated variable, and $\sigma^\prime_i$ and $\mu^\prime_i$
to the dispersion and mean value of the $i$th uncorrelated variable for
the Gaussian distribution.

This method is actually equivalent to perform a generalised $\chi^2$
(such as that in Barreiro, Mart\'\i nez-Gonz\'alez \& Sanz 2001) 
from the original correlated
variables that takes into account the correlations between them.
However, our procedure has some advantages. By uncorrelating
the Gaussian variables, we can have some insight into the significance
of the single detections of non-Gaussianity. In addition, it gives us
the possibility of investigating a further test statistic as follows.
We can construct the probability density function $P(x_i^\prime)$ 
for each of the uncorrelated variables obtained from a large number of
Gaussian simulations. Although the requirement that the variables are
uncorrelated does not imply that they are independent (i.e. that their
joint probability is separable), we may still construct the test
statistic 
\begin{equation}
\eta^2=
\sum_i \ln \left[P(x_i^\prime)\right] ~.
\end{equation}
If the joint probability of ${\mathbf x^\prime}$ is a multidimensional
Gaussian, $\eta^2$ is equivalent to the $\chi^2$ (they coincide except
for an additive constant and a scaling factor).
The construction of a non-Gaussian $\chi^2$ was
already introduced by Ferreira, Magueijo \& G\'orksi (1998) 
to study the normalised
bispectrum of the COBE data. Their definition includes some extra
factors to ensure that $\eta^2$ defaults exactly to the standard $\chi^2$ when
the considered variables follow a multivariate Gaussian distribution.

Unfortunately, we could not estimate this quantity for our
non-Gaussian realisations. This is due to the fact that some of
the values of the variables in the non-Gaussian case lie far beyond a
few $\sigma$'s with 
respect to the distribution of the Gaussian population. This makes it
very difficult to
evaluate the quantity $\eta^2$ for the non-Gaussian distribution.
Instead, we compute the standard $\chi^2$ statistic~(\ref{eqchi2})
and compare its performance to the test described below.

\subsection{The Fisher discriminant function}
\label{fisher}
In the previous tests, we are comparing the level of agreement of a
set of data with some theoretical model. Therefore only information about
this model is included in the test.
However, if we believe that our data are most likely drawn either from 
a hypothesis $H_0$ or from an alternative one $H_1$, it is well known
that an optimal test statistic in the sense of maximum power for 
a given significance level is given by
\begin{equation}
t({\mathbf x})=\frac{f({\mathbf x}|H_0)}{f({\mathbf x}|H_1)}
\end{equation}
which is called the likelihood ratio (Brandt 1992). Unfortunately, in 
order to construct this statistic, we need to know the joint
probabilities $f({\mathbf x}|H_0)$ and $f({\mathbf x}|H_1)$, which are
in general not available.

A simpler possibility is to construct a test statistic using the 
Fisher linear discriminant function (Fisher 1936). 
This is the optimal linear
function of the measured variables for separating the
probabilities $g({\mathbf t}|H_0)$ and $g({\mathbf t}|H_1)$
in the sense of maximum distance between the expected mean values of
$t$ for each model and minimum dispersion of each of them. 
The test statistic is given by
\begin{equation}
t({\mathbf x})=\sum_{i=1}^n a_i x_i = {\mathbf a}^T {\mathbf x}
\end{equation}
with
\begin{eqnarray}
{\mathbf a} &\propto & W^{-1}(\mu_0-\mu_1) \nonumber \\
W_{ij} & = & (V_0+V_1)_{ij}
\end{eqnarray}
where $\mu_k$ and $V_k$ are the expected mean value and covariance matrix of
the measured data if drawn from the hypothesis $H_k$.
Thus $t({\mathbf x})$ is determined up to an additive constant and a
scaling factor.

Actually if $f({\mathbf x}|H_0)$ and $f({\mathbf x}|H_1)$ are both multivariate
Gaussian with common covariance matrices, the Fisher discriminant
function coincides with the likelihood ratio (Cowan 1998).

In our case, the set of variables ${\mathbf x}$ are the fourth
cumulant of the wavelet coefficients at each scale, whereas
the hypotheses $H_0$ and $H_1$ correspond to our simulated data being
drawn from a Gaussian and a non-Gaussian population respectively 
(see \S\ref{results}).
In order to estimate the power of our non-Gaussianity test,
we construct and compare the Fisher discriminant function $t$ for each
Gaussian and non-Gaussian realisation. 
The mean values and covariance matrices of our
variables for both models needed to generate the $W$
matrix are estimated from a large number of simulations.
%?(change the name $t$ for $F$)??

\section{Application to simulated data}
\label{results}
In this section, we apply the wavelet analysis described above to
the detection of non-Gaussianity in simulated CMB maps. 
As our test map, we use the same realisation of CMB anisotropies 
as in HJL. This is a simulated CMB map due to the
Kaiser-Stebbins effect from cosmic strings (Kaiser \& Stebbins 1984). 
The size of the simulation is $6.4 \times 6.4$~deg$^2$ 
with a total of 256$\times$256 pixels of size 1.5 arcminutes.

In order to test the power of our non-Gaussianity test, we add to the
cosmic strings map a Gaussian component that shares the same power
spectrum as the test map. 
The weight of the Gaussian component is controlled by a parameter $b$,
that gives the ratio of the non-Gaussian to Gaussian 
proportions as 1:$b$.
We investigate which is the maximum level of this Gaussian component
(i.e., the largest value of $b$) that can be introduced 
and still detect the underlying non-Gaussianity.

To make more realistic simulations, the test map is convolved with a
Gaussian beam of FHWM=5~arcminutes. Finally, Gaussian white noise
is added with an rms level equal to one-tenth that of the convolved
map. This degree of smoothing and level of noise is typical of what
may be achieved from CMB observations by the Planck satellite of ESA
after foreground removal is performed using the Maximum-Entropy method
(Hobson et al 1998).

As pointed out in the previous section, a single non-Gaussian map was
used to test the power of 
the wavelet analysis in HJL. However, even for the same underlying non-Gaussian
map, the particular value of our statistics depends on the
specific Gaussian component and noise realisation that are being
used. Therefore, the conclusions derived from our test will vary from
one Gaussian realisation to other. To take this into account, we
have analysed a large number (10000) of non-Gaussian test maps 
%in the manner explained in the previous section
that contain the same cosmic strings map
but different Gaussian component and noise realisations.
Ideally, each non-Gaussian map would also contain a different
cosmic strings realisation. Unfortunately, strings maps are difficult to
simulate and very computationally expensive. Therefore, a large number
of stringy maps are not currently available to us and we account only
for the dispersion introduced by the Gaussian components of the
map. Nonetheless, although the particular level of
discrimination achieved by the test may vary
for a different choice of a cosmic strings map, we expect the
general conclusions of this work to remain valid.

Another important point is the choice of the type of wavelet and the 
wavelet basis. We compare the performance of three different 
ways of constructing wavelets: tensor, Mallat and 
{\it \`a trous} algorithm. For the tensor and Mallat algorithms, we also
consider different wavelet bases. In HJL, an
entropy criterion was used for finding the wavelet basis that gives the
single highest detection of non-Gaussianity at a particular scale. In this
paper, however, we perform a joint analysis of the information
provided by all the scales and therefore we cannot apply the same
criterion.
We have tried Haar, Daubechies 4,6,12 and 20, Coiflet 2 and 3
and Symmlet 6 and 8 wavelet bases and find that, using the full analysis,
most of them lead to similar results. In particular, more complex wavelets
(such as Daubechies 12 and 20) tend to discriminate between the models
slightly worse, probably because of a larger number of wavelet
coefficients being discarded due to edge effects. 
Therefore, we present our results for Daubechies~4 and Haar 
wavelets since these are two of the simplest and also
best known orthogonal wavelet bases. Moreover, the Haar basis
functions are antisymmetric and the results of our statistic 
are therefore invariant under rotations of the data (by 90, 180 and
270 degrees).

\subsection{Tensor algorithm}

2D-tensor wavelets are constructed in the manner explained in \S\ref{tensor}.
These are the type of wavelets used in the original work of HJL. 
In this section we present the results for this type of wavelets for
the Daubechies 4 and Haar wavelet bases. We also show the
discriminating power of the Fisher discriminant function versus a $\chi^2$ 
analysis and address the orientation sensitivity problem.

\subsubsection{Daubechies 4}
\begin{figure}
\centerline{\epsfig{file=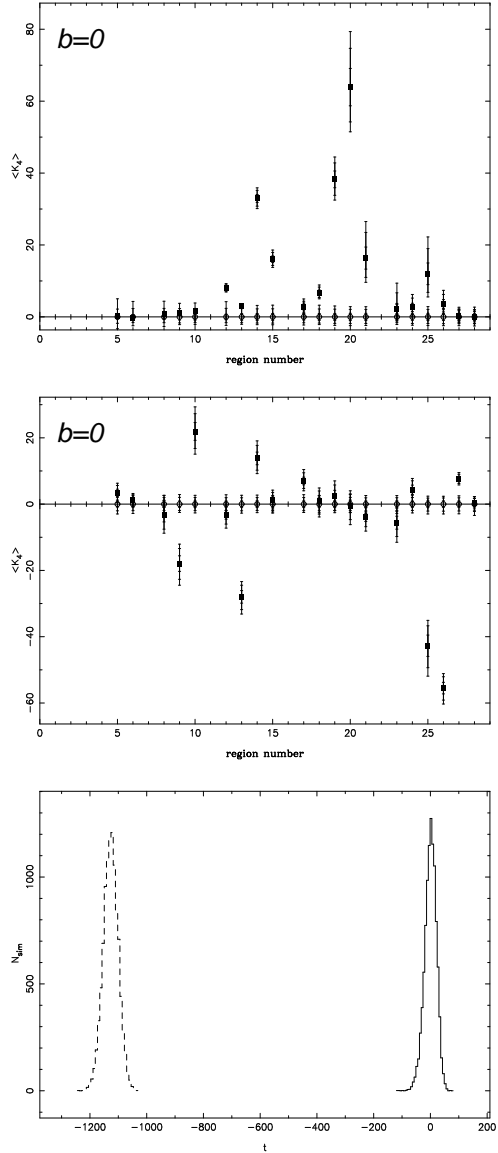,width=6.5cm}}
\caption{Top panel shows the fourth cumulant spectra obtained from
10000 Gaussian and non-Gaussian simulations for $b=0$, using tensor
Daubechies 4 wavelets. The corresponding uncorrelated variables are
given in the middle panel. The bottom panel shows the histogram of the
Fisher discriminant obtained from the Gaussian (solid line) and
non-Gaussian (dashed line) simulations (see text for details).}
\label{b10_d4_rot0}
\end{figure}
The top panel of Fig.~\ref{b10_d4_rot0} shows
the fourth cumulant spectrum for the test map with non-Gaussian:Gaussian
proportions (1:0) using Daubechies 4 and the original orientation of
the data. Solid squares and open circles correspond to the 
non-Gaussian and Gaussian values respectively. The Gaussian values
have been normalised to zero mean value and unit dispersion to allow
for a straightforward reading in units of the Gaussian dispersion 
of the distance between the two models. The error bars show the 68, 95
and 99 per cent confidence levels of the Gaussian and non-Gaussian 
distributions derived from 10000 simulations in each case.
The scale corresponding to the region number is given in
Table~\ref{regions_tensor}; a higher region number corresponds, in
general, to 
smaller scales. We see that there are several large detections of
non-Gaussianity, the highest single detection being $\sim 60 \sigma$
in region 20. This corresponds to scales of a few arcminutes ($\simeq
6-12'$) which is expected since the characteristic signature of the
Kaiser-Stebbins effect shows at small scales. 
As an illustration, we also show in the middle panel the corresponding 
rotated uncorrelated variables. These are obtained as explained in
\S\ref{chitest}.  
\begin{table}
\caption[]{Correspondence of region number with scale for tensor
wavelets. The scale is calculated as $1.5'(2^{J-j_1}\times
2^{J-j_2})$, where $2^J\times 2^J$ is the total number of pixels in the map.
The fifth and sixth columns give the total number of coefficients in
the region and how many of those coefficients are used to
calculate the fourth cumulant of their distribution when using
Daubechies 4. For the Haar wavelet, there is not need of correction
for boundary effects and all coefficients are used. Finally, the last
column indicates the number of the corresponding region for Mallat
wavelets, for those regions that are common for both types of
wavelets. For symmetry, wavelet coefficients in domains with 
$j_2 > j_1$ are included in the same region as those with scales 
$(j_2,j_1)$. Following HJL, wavelet coefficients with $j_i=0$ are not
included in the analysis.}
\label{regions_tensor}
\begin{tabular}{|c|c|c|c|c|c|c|}
\hline
region & $j_1$ & $j_2$ & $\sim$scale($'$) & no.coeff. & used (D4) & Mallat \\
\hline
1      & 1     & 1     & $192 \times 192$    & 4           & 0      & 3 \\
2      & 2     & 1     & $96 \times 192$      & 16          & 0      &   \\
3      & 2     & 2     & $96 \times 96$      & 16          & 0      & 6 \\
4      & 3     & 1     & $48 \times 192$      & 32          & 0      &   \\
5      & 3     & 2     & $48 \times 96$      & 64          & 10     &   \\
6      & 3     & 3     & $48 \times 48$      & 64          & 25     & 9 \\
7      & 4     & 1     & $24 \times 192$      & 64          & 0      &   \\
8      & 4     & 2     & $24 \times 96$      & 128         & 26     &   \\
9      & 4     & 3     & $24 \times 48$    & 256         & 130    &   \\
10     & 4     & 4     & $24 \times 24$   & 256         & 169    & 12\\
11     & 5     & 1     & $12 \times 192$     & 128         & 0      &   \\  
12     & 5     & 2     & $12 \times  96$   & 256         & 58     &   \\  
13     & 5     & 3     & $12 \times  48$   & 512         & 290    &   \\  
14     & 5     & 4     & $12 \times  24$   & 1024        & 754    &   \\  
15     & 5     & 5     & $12 \times 12 $   & 1024        & 841    & 15\\  
16     & 6     & 1     & $6 \times  192$   & 256         & 0      &   \\  
17     & 6     & 2     & $6  \times 96$   & 512         & 122    &   \\  
18     & 6     & 3     & $6  \times 48$   & 1024        & 610    &   \\  
19     & 6     & 4     & $6 \times  24$   & 2048        & 1586   &   \\  
20     & 6     & 5     & $6 \times  12$   & 4096        & 3538   &   \\  
21     & 6     & 6     & $6 \times  6 $   & 4096        & 3721   & 18\\  
22     & 7     & 1     & $3 \times 192$    & 512         & 0      &   \\  
23     & 7     & 2     & $3 \times 96 $   & 1024        & 252    &   \\  
24     & 7     & 3     & $3\times  48 $   & 2048        & 1260   &   \\  
25     & 7     & 4     & $3\times  24 $   & 4096        & 3276   &   \\  
26     & 7     & 5     & $3\times  12 $   & 8192        & 7308   &   \\  
27     & 7     & 6     & $3\times  6 $   & 16384       & 15372  &   \\  
28     & 7     & 7     & $3\times  3$    & 16384       & 15876  & 21\\  
\hline
\end{tabular}
\end{table}
\begin{table*}
\caption[]{Discriminating power of the non-Gaussianity test for
different wavelets}
\label{power}
\begin{tabular}{|c|c|c|c|c|c|c|c|}
\hline
& \multicolumn{3}{c}{Tensor} & \multicolumn{3}{c}{Mallat} 
& {\it \`A trous}  \\
& \multicolumn{2}{c}{Daubechies 4}  & Haar & 
\multicolumn{2}{c}{Daubechies 4} & Haar & \\
&$0^\circ$ & $180^\circ$ & &$0^\circ$ & $180^\circ$ & & \\
\hline
$b=0$ & 1.00 & 1.00 & 1.00 & 1.00 & 1.00 & 1.00 & 1.00 \\
$b=1$ & 1.00 & 1.00 & 1.00 & 1.00 & 1.00 & 1.00 & 1.00 \\
$b=2$ & 0.517& 0.798& 0.619& 0.974& 0.996& 0.995& 0.991 \\
$b=3$ & 0.038& 0.091& 0.037& 0.165& 0.330& 0.252& 0.187 \\
%noisy & 0.999& 1.00 & 1.00 & 1.00 & 1.00 & 1.00 & 1.00 \\
\hline
\end{tabular}
\end{table*}

To evaluate in a more quantitative way the discriminating power of our
non-Gaussianity test, we have obtained the Fisher discriminant function
$t$ as explained in \S~\ref{fisher}. This is shown in the bottom panel of
Fig.~\ref{b10_d4_rot0}. The solid and dashed lines correspond to the
histogram of $t$ for the Gaussian
and non-Gaussian simulations respectively. As expected from the
difference in the fourth cumulant spectrum, the curves
are clearly separated. Indeed the power of the test to discriminate
between both models at the 0.01 significance level is 1.
\begin{figure*}
\centerline{\epsfig{file=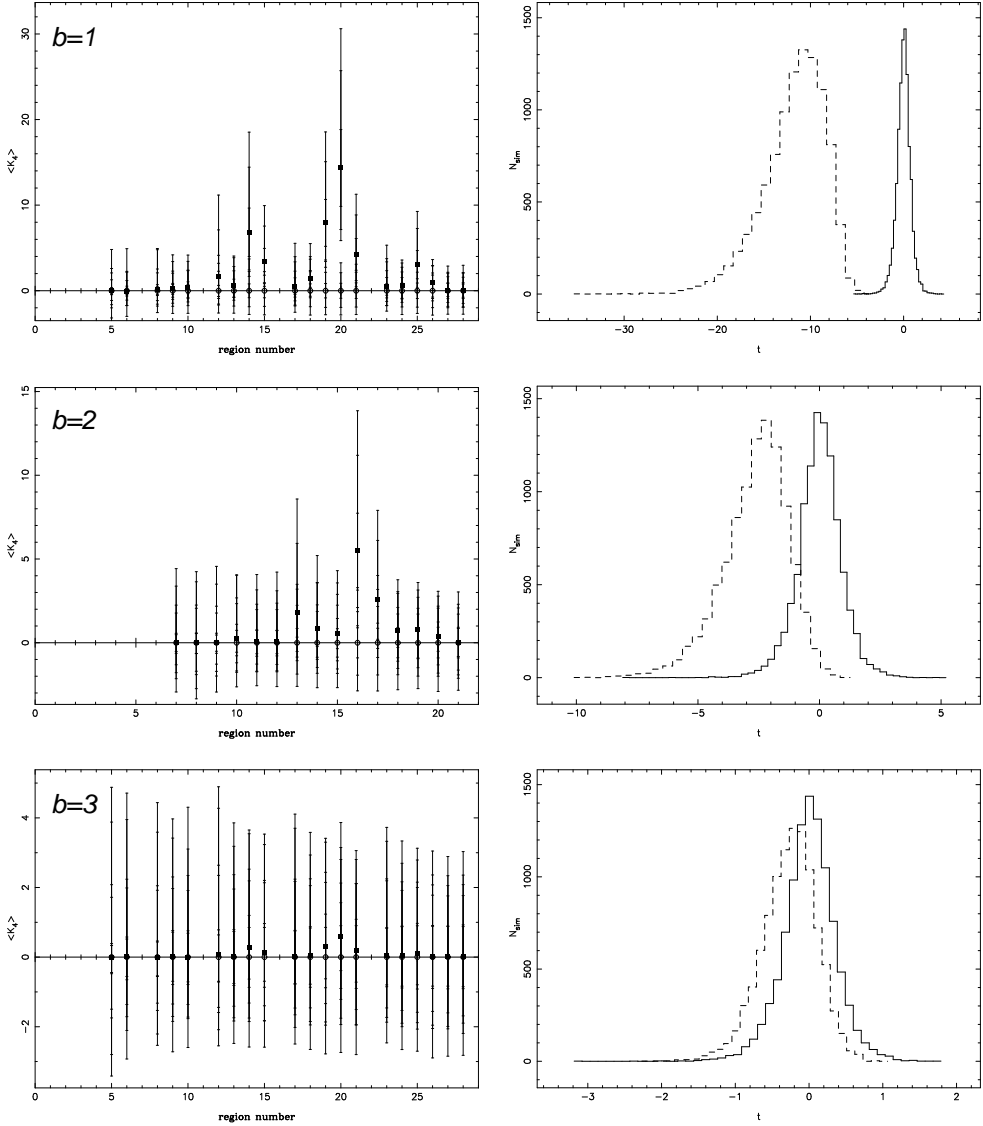,width=13cm}}
\caption{Fourth cumulant spectra and Fisher discriminant for $b$=1,2,3
using the tensor Daubechies 4 wavelet. Open circles/solid lines and
solid squares/dashed lines
correspond to the Gaussian and non-Gaussian models respectively. }
\label{b123_d4_rot0}
\end{figure*}

In order to assess the amount of Gaussian confusion that can be added
to our test map and still detect the underlying non-Gaussianity, we
increase the value of $b$ until the models cannot be
separated. Fig.~\ref{b123_d4_rot0} shows the fourth cumulant spectrum and
Fisher discriminant function for $b=1,2,3$. 
For non-Gaussian:Gaussian proportions of 1:1, the presence of the
non-Gaussian signature is still clear at region 20,
although the level of the detection has fallen appreciably. 
This translates into much closer histograms for the Fisher discriminant
function but the power of the test at the 0.01 significance level
remains 1. When the level of Gaussian confusion is increased to $b=2$, 
the fourth cumulant spectrum of both models shows no clear detections.
Although some differences are seen in the estimated mean 
values of both
populations, the error bars clearly overlap. However, when all the
information is taken into account simultaneously using the Fisher
discriminant, the models can be partially separated. We find that the
power of the test is $p=0.517$.
This means that we are able to detect the non-Gaussian signal in
approximately half of our simulated test maps.
For $b=3$, the non-Gaussianity signal is completely obscured by the
Gaussian component, and the test fails to discriminate between both
models ($p=0.038$). Table~\ref{power} summarises the values of $p$ for
the different cases considered.

\begin{figure}
\centerline{\epsfig{file=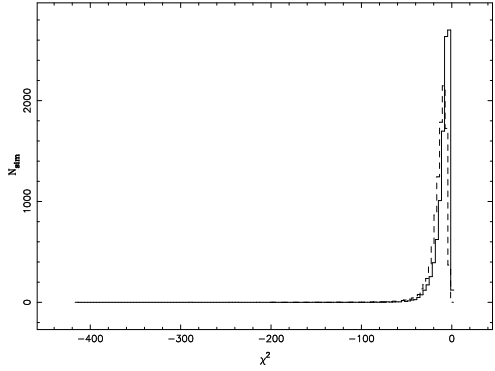,width=6.5cm}}
\caption{Histogram of $\chi^2$ values as derived from the fourth
cumulant spectra of the Gaussian (solid line)
and non-Gaussian (dashed lines) simulations 
for $b=2$ when using tensor Daubechies 4 wavelets. The $\chi^2$ values
were obtained as explained in \S\ref{chitest}.}
\label{chi2}
\end{figure}

As pointed out in the previous section, the Fisher
function is the optimal linear discriminant in the sense of 
maximising the power of the test.
It is interesting to compare how the standard $\chi^2$ test
performs in comparison with the previous results.
To illustrate this point, we plot the
histogram for the $\chi^2$ values obtained from the Gaussian and non-Gaussian
simulations for the case $b=2$ (see Fig.~\ref{chi2}). We see that in
this case the curves for both models completely overlap and the power
of the test drops to $p=0.017$ to be compared with $p=0.517$ in the
case of the Fisher discriminant function. Indeed, we find that the Fisher
discriminant function clearly outperforms the power of the $\chi^2$
test in most cases. Only in those cases where the differences between
the models are very large (for instance when $b=0$), do both tests give
comparable results.

\begin{figure}
\centerline{\epsfig{file=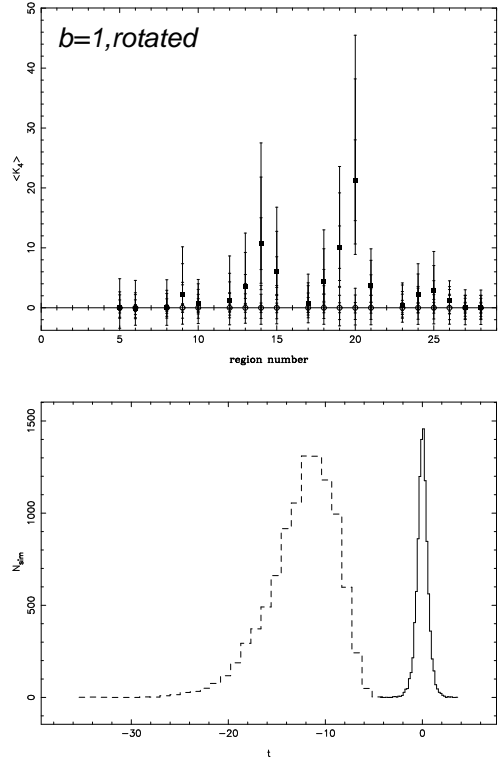,width=6.5cm}}
\caption{Fourth cumulant spectrum and histogram of Fisher
discriminant for the Gaussian and non-Gaussian simulations ($b=1$)
rotated by 180 degrees with respect to the original
orientation. Tensor Daubechies 4 wavelets were used.}
\label{b11_d4_rot180}
\end{figure}

Another point that needs to be addressed is the orientation
sensitivity of the test due to the assymetry of the wavelet basis
functions. 
Because of the homogeneity and isotropy of the CMB temperature
field, different orientations of the data should be statistically
equivalent. However, if we look at one single realisation the results
will vary when rotating the data. This is also what is happening in
our case, since we are using the same underlying
cosmic strings realisation for all the non-Gaussian maps.
Therefore, repeating the former analysis for a different orientation
of our simulated maps will give us an insight on the importance of this effect
when analysing a particular set of real data. Thus, we have performed
the analysis for the same Gaussian and non-Gaussian maps
%In order to do this, we have repeated the former analysis
%for the same Gaussian and non-Gaussian maps 
rotated 180 degrees with
respect to the original orientation. The discriminating power of the
test for these cases are given in Table~\ref{power}. These numbers are
typical of the range of variation of the power of the test for
different rotations. For illustration, we show the case for $b=1$ in
Fig.~\ref{b11_d4_rot180}. We see that the single detection of non-Gaussianity
at region 20 is larger than that of the original orientation
(see middle panels of Fig.\ref{b123_d4_rot0}). However, when all the
information is taken into account, the overlapping between both models
is quite similar. For larger levels of confusion ($b=2,3$), the Fisher
discriminant function is sensitive even to small differences in the
detection level of non-Gaussianity, leading to a better separation of
the models for the rotated data. In particular, $p=0.798$ for $b=2$
versus $p=0.517$ for the original orientation of the maps.
Therefore, our non-Gaussianity test is sensitive to the orientation of 
the data. In order to deal with this problem, we need to use
symmetric (or anti-symmetric) wavelet basis functions. For orthogonal
wavelets, only the Haar wavelet fulfils this requirement. Thus,
in the next subsection we repeat our analysis using this wavelet.
A second possibility is to use a symmetric non-orthogonal wavelet, such
as the one constructed with the {\it \`a trous} algorithm (see \S\ref{atrous}).

\subsubsection{Haar}

As already pointed out, using the Haar wavelet solves the problem of
the orientation sensitivity of our non-Gaussianity test. Indeed, with this
wavelet basis, our test is invariant under
rotations of 90, 180 and 270 degrees of the original data.
Moreover, there are no edge effects when using the Haar wavelet, and
therefore all the wavelet coefficients can be used in the analysis.
Haar is also the simplest orthogonal wavelet to implement.
\begin{figure*}
\centerline{\epsfig{file=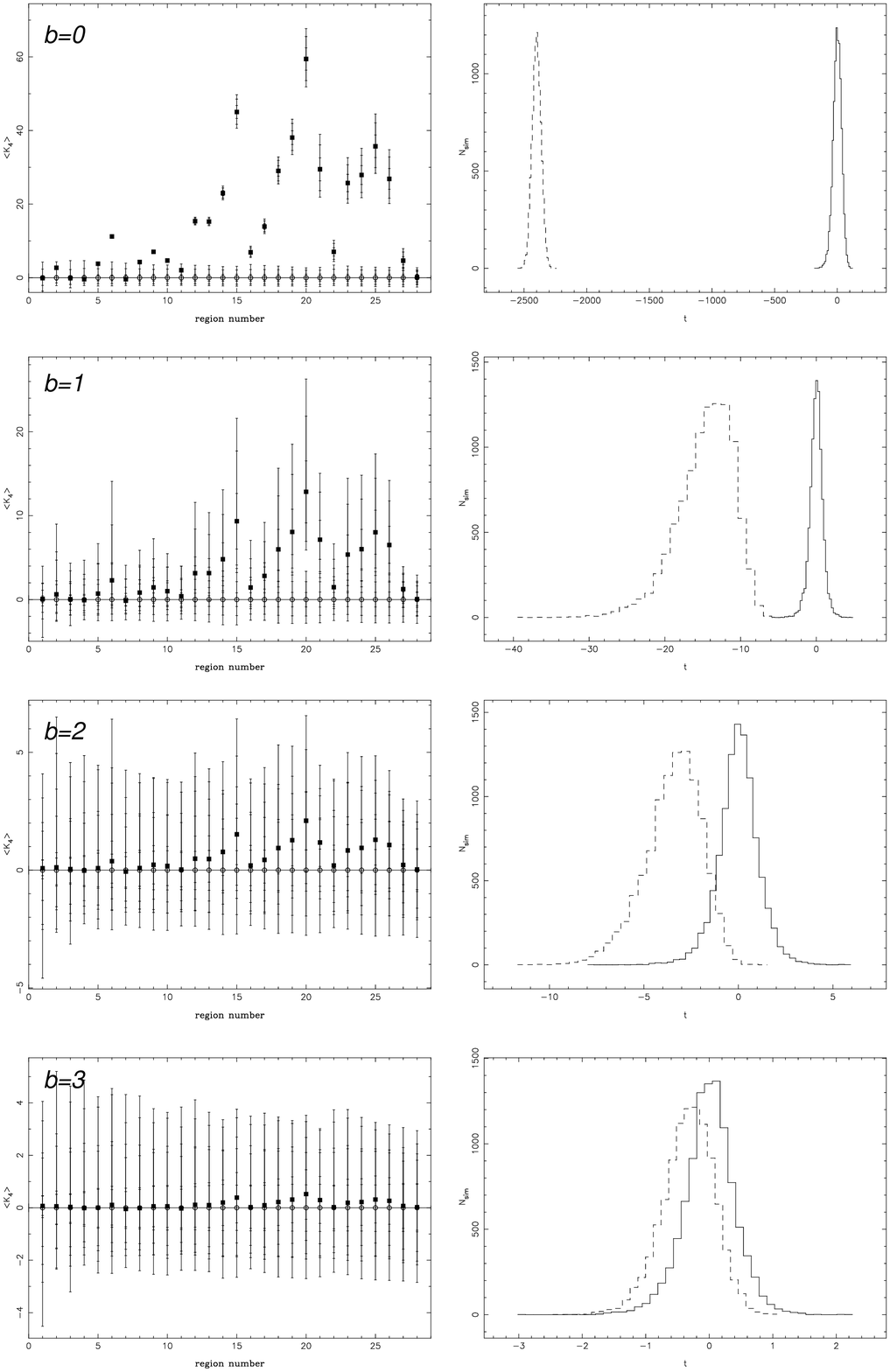,width=13.0cm}}
\caption{Fourth cumulant spectrum and Fisher discriminant for
different values of $b$ using tensor Haar wavelets.}
\label{b0123_h}
\end{figure*}
In Fig.~\ref{b0123_h} we plot the fourth cumulant spectra and the histograms
of the Fisher discriminant for $b=0,1,2,3$. We see that the Gaussian
and non-Gaussian models are successfully 
separated for $b=0$ and $b=1$, with a discriminating power of $1$
(see Table~\ref{power}). When the level of Gaussian confusion is
increased the models start to overlap ($p=0.619$ for $b=2$) and
finally mix completely ($p=0.037$ for $b=3$).
These results are very similar to those obtained using Daubechies 4
(see Figs.~\ref{b10_d4_rot0} and~\ref{b123_d4_rot0}).
We note that the discriminating power of the test
for $b=2$ in this case lies in between the values
obtained for the two different rotations of the data when the analysis
is performed with Daubechies 4. 

Therefore the Haar wavelet gives similar results to the Daubechies 4
case, but it also has the property of rotational invariance.

\subsection{Mallat algorithm}
In this section we repeat the previous analysis for Daubechies 4 and
Haar but using 2-dimensional wavelets constructed as explained in
\S\ref{mallat}. As already said, for this type of wavelet,
there are three different kind of coefficients: vertical, horizontal
and diagonal. 
Horizontal and vertical details should be statistically equivalent,
but we obtain different levels of detection because we are looking at
a particular realisation (the same underlying
non-Gaussian map), as it would happen for a set of real data.
The numbering of the regions is as follows: regions
1,2,3 correspond to vertical, horizontal and diagonal details
respectively for the largest scale ($j=1$); regions 4,5,6 give the
three details in the same order for the next scale ($j=2$) and so on
(see Table~\ref{regions_mallat}).

\begin{table}
\caption[]{Correspondence of region number with scale for Mallat
algorithm. The scale is calculated as $1.5'(2^{J-j})$.
The total number of coefficients and those used in the analysis of each
region when using Daubechies 4 are given in the fifth and sixth
columns. All the coefficients are used for the Haar wavelet.
The corresponding region number for tensor wavelets of those regions
common to both kind of wavelets are given in the last column.}
\label{regions_mallat}
\begin{tabular}{|c|c|c|c|c|c|c|}
\hline
region & $j$ & detail & $\sim$scale($'$) & no.coeff. & used (D4) & Tensor \\
\hline
1  & 1 & V & 192 & 4     & 0     &    \\
2  & 1 & H & 192 & 4     & 0     &    \\
3  & 1 & D & 192 & 4     & 0     & 1  \\
4  & 2 & V & 96 & 16    & 0     &    \\
5  & 2 & H & 96 & 16    & 0     &    \\
6  & 2 & D & 96 & 16    & 0     & 3  \\
7  & 3 & V & 48 & 64    & 25    &    \\
8  & 3 & H & 48 & 64    & 25    &    \\
9  & 3 & D & 48 & 64    & 25    & 6  \\
10 & 4 & V & 24 & 256   & 169   &    \\
11 & 4 & H & 24 & 256   & 169   &    \\
12 & 4 & D & 24 & 256   & 169   & 10 \\ 
13 & 5 & V & 12 & 1024  & 841   &    \\
14 & 5 & H & 12 & 1024  & 841   &    \\
15 & 5 & D & 12 & 1024  & 841   & 15 \\ 
16 & 6 & V & 6 & 4096  & 3721  &    \\
17 & 6 & H & 6 & 4096  & 3721  &    \\
18 & 6 & D & 6 & 4096  & 3721  & 21 \\
19 & 7 & V & 3 & 16384 & 15876 &    \\
20 & 7 & H & 3 & 16384 & 15876 &    \\
21 & 7 & D & 3 & 16384 & 15876 & 28 \\
\hline
\end{tabular}
\end{table}

\subsubsection{Daubechies 4}
\begin{figure*}
\centerline{\epsfig{file=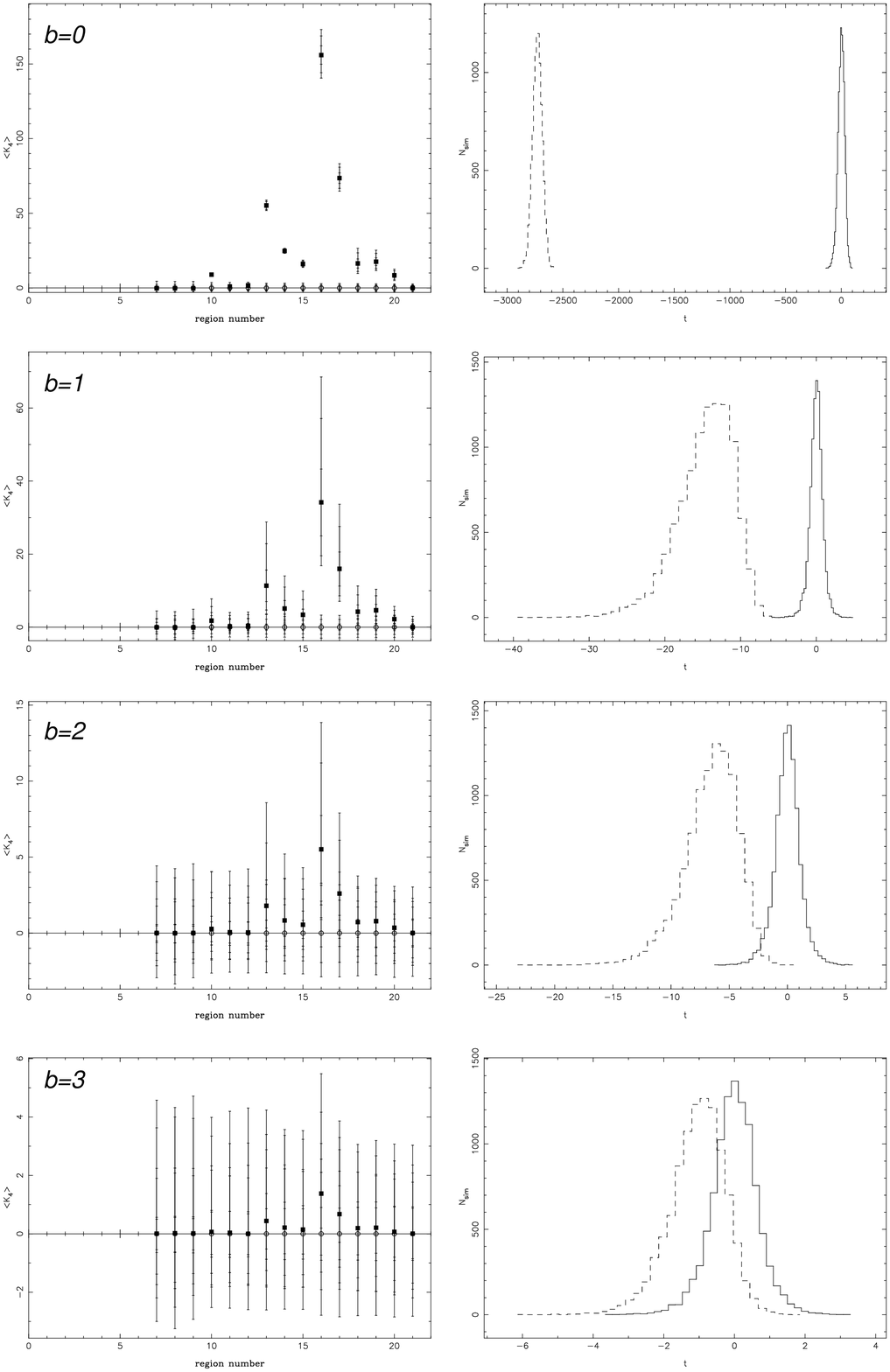,width=13.0cm}}
\caption{Fourth cumulant spectrum and Fisher discriminant for
different values of $b$ using Mallat Daubechies 4 wavelets.}
\label{b0123_d4_rot0_mra}
\end{figure*}

Figure~\ref{b0123_d4_rot0_mra} shows the fourth cumulant spectra for
different non-Gaussian:Gaussian proportions ($b=0,1,2,3$) as well as
the histograms of the Fisher discriminant function for the Gaussian
and non-Gaussian models. The values of the power of the test are summarised
in Table~\ref{power}. Again we see how the models begin to overlap
as the level of the Gaussian component is increased. However, the
test is more powerful at discriminating between the models 
than the Daubechies 4 tensor case (see Figs.~\ref{b10_d4_rot0}
and~\ref{b123_d4_rot0}). There are a larger number of clear 
detections which also have a higher amplitude. In particular,
the highest detection is found at region 16, 
which corresponds to the vertical detail coefficients
at scales of $\sim 6'$. The higher discriminating power of these
wavelets becomes especially clear for the case $b=2$, where we find
a value of $p=0.974$ versus $p=0.519$ for the tensor wavelets.
%we find a value of $p=0.974$ for the case $b=2$ 
%versus $p=0.519$ for the tensor wavelets.
We note that some of the regions are
common for both the tensor and Mallat wavelets (see
Table~\ref{regions_mallat}). However, the highest detections of
non-Gaussianity do not appear in these regions.
In particular, for the Mallat algorithm,
the non-Gaussian detection in the horizontal and vertical details is
usually larger than that of the diagonal wavelet
coefficients for the same scale.
This may be due to the type of non-Gaussianity present in our strings
map, which contain sharp features that are more easily detected on the
horizontal and vertical details.
Analogously, for the tensor case, the largest detections are obtained
in regions with slightly different values of $j_1$ and $j_2$, since they seem
to identify better the features of the long strings.

The corresponding values for a different orientation of the data
(rotated 180 degrees with respect to the original maps) are given in
Table~\ref{power}. As for the tensor case, the single detections of
non-Gaussianity can appreciably vary for different rotations.
However, when all the information is taken into account in
the Fisher discriminant, they lead to similar results, at least
for the cases of moderate levels of confusion.

\subsubsection{Haar}
\begin{figure*}
\centerline{\epsfig{file=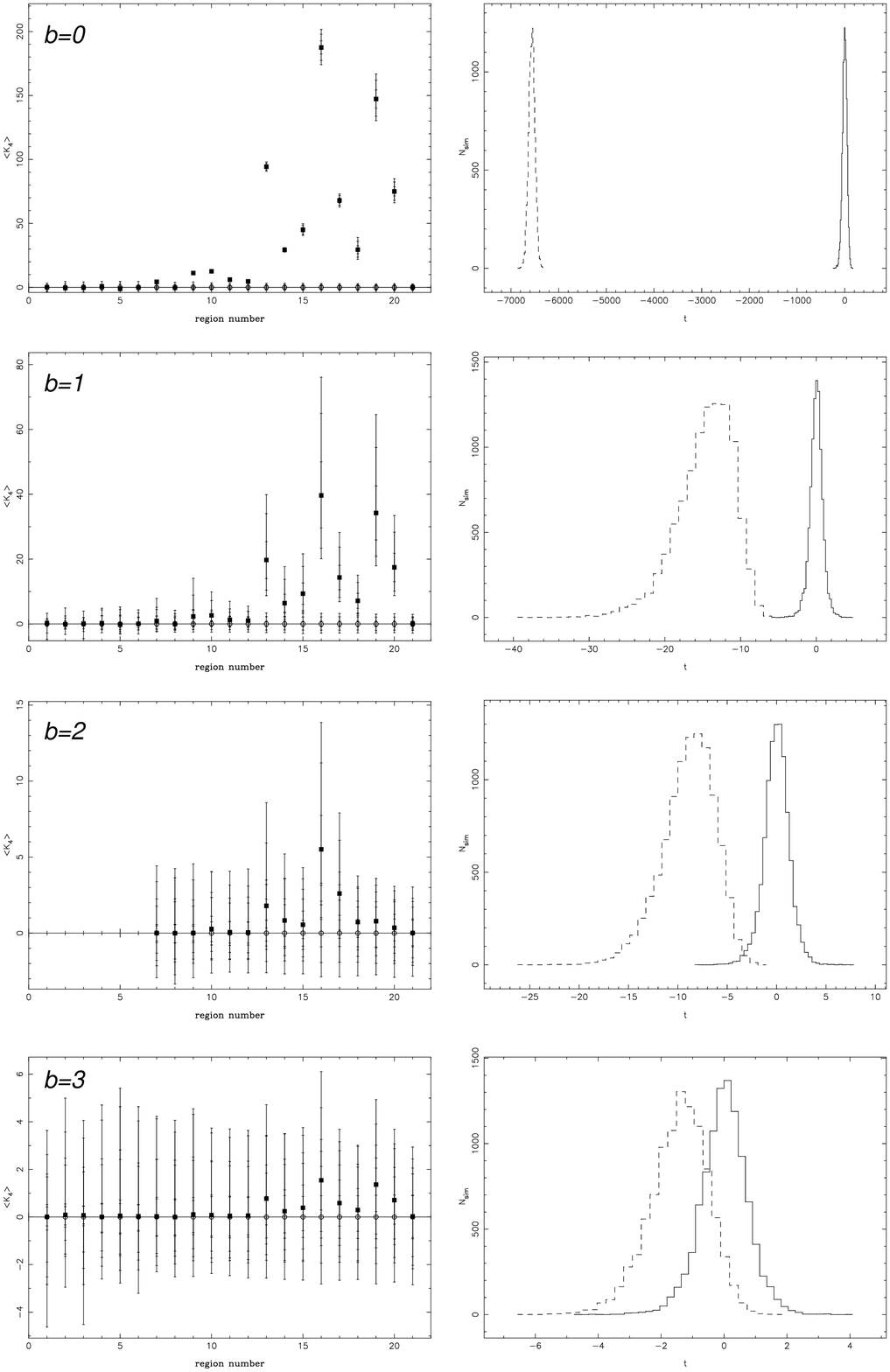,width=13.0cm}}
\caption{Fourth cumulant spectrum and Fisher discriminant for
$b=0,1,2,3$ for Mallat Haar wavelets.}
\label{b0123_h_mra}
\end{figure*}

In order to address the orientation sensitivity problem, we repeat 
the former analysis again using the Haar wavelet.
The fourth cumulant spectra and Fisher discriminant function for $b=0,1,2,3$
are given in Fig.~\ref{b0123_h_mra}. Although the amplitude of the
single non-Gaussian detections is different from the Daubechies 4
case (see Fig.~\ref{b0123_d4_rot0_mra}), the discriminating power of the
test is actually very similar. In fact, the values of
$p$ for the Haar wavelet, lies between the values for the two
considered orientations of the data. We find again that the Mallat
wavelet works better discriminating between both models than the
tensor construction when the same wavelet basis is used (see
Table~\ref{power}).

\subsection{{\it \`A trous} algorithm}
\begin{figure*}
\centerline{\epsfig{file=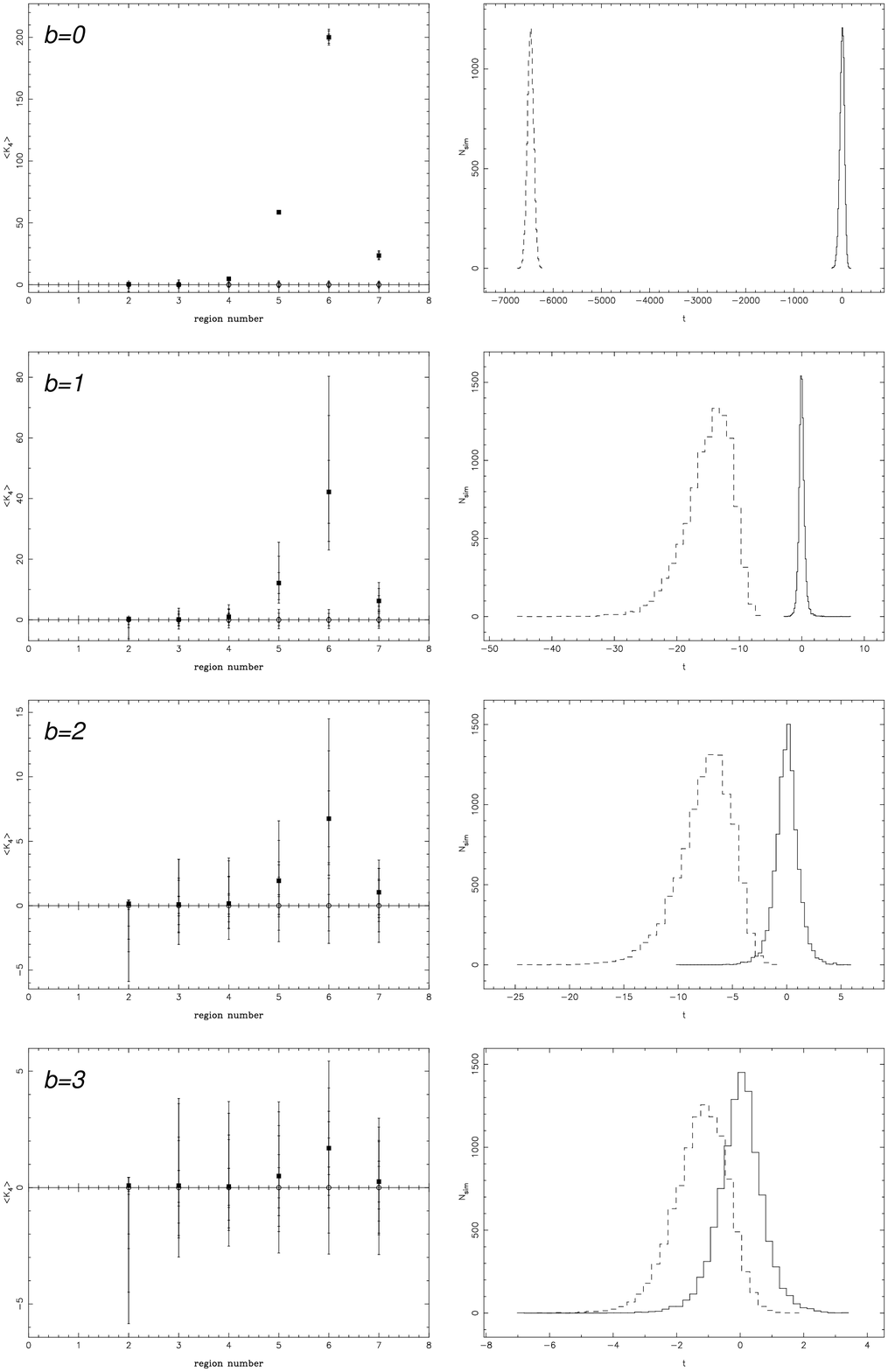,width=13cm}}
\caption{The fourth cumulant spectra and the corresponding histograms
of the Fisher discriminant are given for various values of $b$
obtained when using the {\it \`a trous} algorithm.}
\label{b0123_atrous}
\end{figure*}

\begin{figure*}
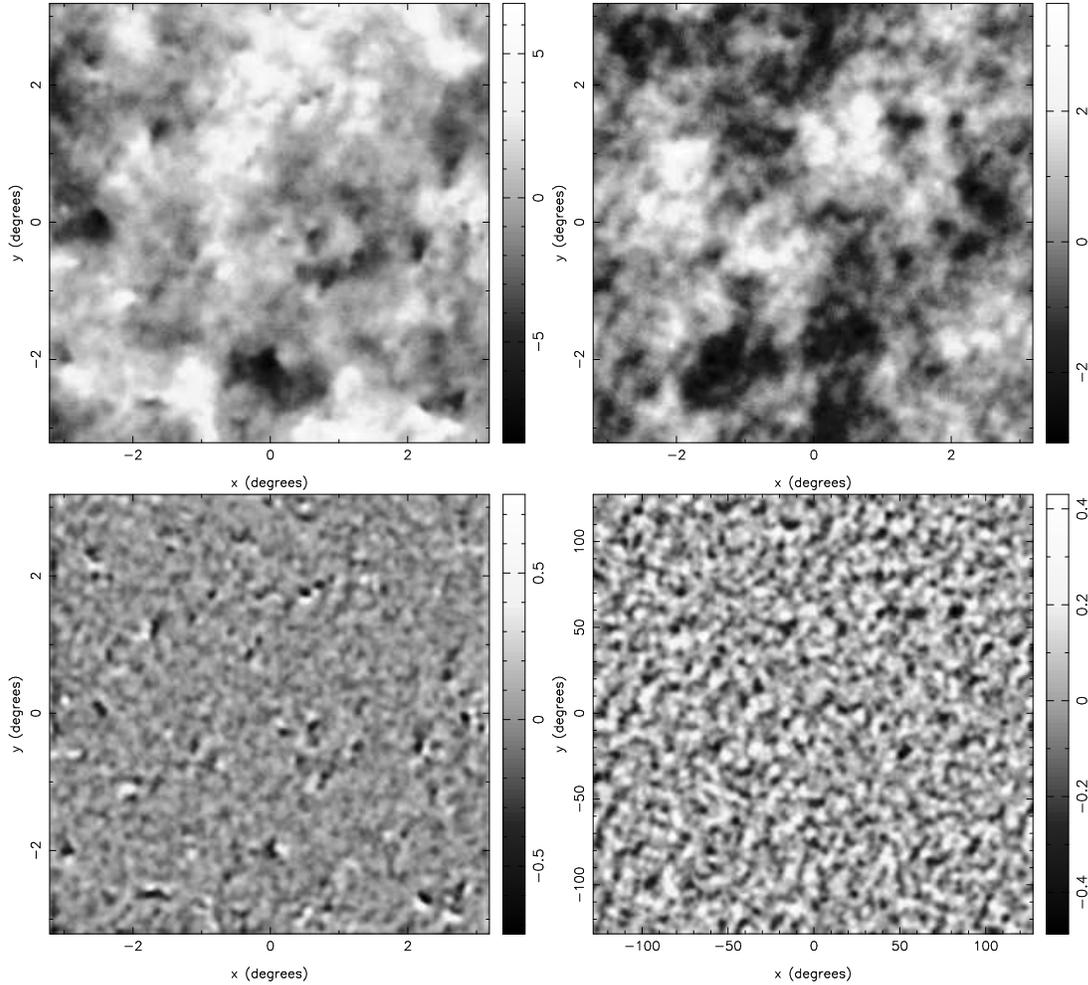

\epsfig{file=mapng_bw.ps,width=6.5cm,angle=-90}
\epsfig{file=mapg_bw.ps,width=6.5cm,angle=-90}
\epsfig{file=mapng_scale6_bw.ps,width=6.5cm,angle=-90}
\epsfig{file=mapg_scale6_bw.ps,width=6.5cm,angle=-90}
\caption{One of our non-Gaussian simulated maps with $b=1$ (top left) and an
equivalent Gaussian realisation (top right). The bottom panels show
the corresponding wavelet coefficients at scale $j=6$ for the {\`a
trous} algorithm.}
\label{plotatrous}
\end{figure*}

In this section we compute the results for the same test but now 
based on the {\it \`a trous} algorithm. As explained in
\S\ref{atrous}, this is a redundant
non-orthogonal wavelet basis, that contains the same number of wavelet
coefficients at each scale as the original image. 
Since we have chosen a symmetric mask, the {\`a trous} wavelet transform 
is invariant under rotations of the data of 90, 180 and
270 degrees with respect to the original orientation.
\begin{table}
\caption[]{Correspondence of region number with scale for the {\it \`a
trous} algorithm. The scale is estimated as 
$1.5'(2^{J-j})$
The last two columns give the total number and the
number of coefficients used to calculate the fourth cumulant at each region}
\label{regions_atrous}
\begin{tabular}{|c|c|c|c|c|}
\hline
region & $j$ & $\sim$scale($'$) & no.coeff. & used \\
\hline
2 & 2 & 96 & 65536 & 16    \\
3 & 3 & 48 & 65536 & 17424 \\
4 & 4 & 24 & 65536 & 38416 \\
5 & 5 & 12 & 65536 & 51984 \\
6 & 6 & 6 & 65536 & 59536 \\
7 & 7 & 3 & 65536 & 63504 \\
\hline
\end{tabular}
\end{table}
Fig.~\ref{b0123_atrous} shows the results for the fourth cumulant spectra and
the Fisher discriminant when the {\it \`a trous} algorithm is used for
$b=0,1,2,3$. The region
number coincides with the $j$ scale of the wavelet. Therefore, higher
region numbers correspond to smaller scales (see
Table~\ref{regions_atrous}). We see that, the highest 
single detection occurs at region 6, that corresponds to scales of
$\sim 6'$. This coincides with the scale where the maximum
detection was found for the former studied wavelets.
As an illustration, Fig.~\ref{plotatrous} also shows one of our
non-Gaussian (top left panel) and Gaussian (top right panel)
temperature maps for $b=1$ and the corresponding wavelet coefficients
(bottom panels) at scale $j=6$, where the maximum detection is
found. We see clear differences in the structure
of the wavelet coefficients maps, which are however not appreciated
in the maps themselves. In fact, the wavelet transform succeeds 
on picking up the butterfly-like signature of the Kaiser-Stebbins
effect which is seen in the wavelet coefficients of the
non-Gaussian map.
In spite of having only six different scales, that contain
repeated information (since each region is comprised of the same number of
wavelet coefficients as the original image) the separation of the
models is better than that obtained with the tensor wavelets and
comparable to the Mallat algorithm case (see Table~\ref{power}).

\bigskip
In HJL, the case where Gaussian white noise is the main contaminant 
to the CMB map was also considered. 
This has been traditionally the greatest obstacle to the
detection of non-Gaussianity in the CMB, and this is still the case
for the 4-yr COBE data. However, current and future experiments are
expected to produce data with a much higher signal to noise ratio and
therefore this case offers less interest. Due to this fact and for the
sake of brevity, we do no include this case in the present
work. Nonetheless, we have
repeated our analysis for maps where the rms level of instrumental
noise is equal to that of the convolved CMB map and find that all the
considered wavelets perform very well. In fact the two models are clearly
discriminated being the power of the test $p=1$. We also find
that the highest detection move towards larger scales. This
is expected since the noise is most contributing at small scales,
therefore obscuring the non-Gaussian signal.

\section{Discussion and conclusions}
\label{concl}
We have investigated the performance of wavelet techniques to detect
and characterise non-Gaussianity in the CMB, extending the work of
Hobson, Jones \& Lasenby (1999).
Our test map consists of a realisation 
of the Kaiser-Stebbins effect due to cosmic strings of size $6.4$
square degrees and pixel of 1.5 arcminutes. In addition, a
Gaussian component with the same power spectrum as the original signal
has been superposed with Gaussian:non-Gaussian proportions
$1:b$ and the resulting signal convolved with a Gaussian beam of
$FWHM=5$ arcminutes.
Gaussian white noise was also added with an rms equal to
one-tenth that of the smoothed signal. A large number of
non-Gaussian maps (10000) were generated, containing the same
underlying Kaiser-Stebbins map and different realisations of the
Gaussian component and noise. This allows us to account for the
dispersion due to the Gaussian component and noise although the
results will depend on the particular non-Gaussian underlying cosmic
string map that is being used. Ideally, different non-Gaussian components
should have been used for each map but generating cosmic strings maps is
very computationally expensive and a large number of them were
not available to us. However, these maps are still a good test for
the method presented and, even if the particular level of 
discrimination may vary from realisation to realisation, 
we expect the general conclusions of this work to remain valid.
The same number of Gaussian simulations were generated, simply by
randomising the phases of the non-Gaussian maps. This method ensures
that Gaussian and non-Gaussian realisations 
share the same power spectrum and thus any difference
found between both models will be due to the non-Gaussian character of
the signal and not to differences in the power spectrum.

The non-Gaussian test is as follows. First, we have obtained the 
wavelet transform of our non-Gaussian and
Gaussian maps. We have then computed the fourth cumulant at each
wavelet scale for each of the maps. All this
information has been combined 
by computing the Fisher discriminant function for each
simulated map and, finally, the distribution of this quantity has been
compared for the Gaussian and non-Gaussian models.
Indeed, the choice of a good test statistic that combines all the 
information from the different wavelet scales is crucial.
The Fisher discriminant function 
is the optimal linear function of the measured variables
in the sense of maximum power.
In fact, we have found that, for most of our cases, the Fisher 
discriminant function clearly outperforms a generalised $\chi^2$ test.

We have also compared the performance of different wavelet
transforms. Wavelets constructed using the Mallat algorithm have been 
shown to perform better than the 2D-tensor wavelets (used in the
original work of HJL). In particular, using the Mallat algorithm, the stringy
and inflationary models can be discriminated at a very high confidence
level even when the Gaussian confusion has a rms twice as large as the one of
the cosmic strings map ($b=2$, $p=0.995$ for Haar). However, the
discriminating power of the test 
is clearly lower for the same case when tensor
wavelets are used ($p=0.619$). 
In addition, we have tried different orthogonal bases for both
decomposition schemes and find very similar
results, providing all the information is combined into the Fisher
discriminant function. In general, we have found that simpler wavelet
bases perform slightly better than more complex ones.
Thus we have used the Daubechies 4 and Haar wavelets for our work.

We have also investigated the performance of the {\`a trous}
algorithm, a non-orthogonal wavelet transform. In spite of being
redundant (each scale contains as many coefficients as the original
image), its discriminating power is very similar to that of the Mallat
algorithm. In particular, we find $p=0.991$ for $b=2$.

We have also addressed the problem of orthogonal wavelet transforms being
sensitive to the orientation of the data, which is a consequence of the
assymetry of the wavelet basis functions. To avoid this problem, we have used
the Haar wavelet functions, which are antisymmetric. This means that
the non-Gaussianity test is invariant under rotations of the data when
the Haar wavelet is used. Another possibility is using a
symmetric non-orthogonal wavelet such as the {\`a trous} algorithm.
We have also presented results for Daubechies 4 for two different 
orientations of the data to give an
insight of the sensitivity of the test when the original map is
rotated.

In addition, wavelet techniques are useful to characterise the
angular scale at which the non-Gaussian signal occurs. We have found
the maximum non-Gaussian signal at scales $\sim 6$~arcminutes for the
different considered transforms. Finally, we note that our test is
not making use of the localisation property of wavelets. However, we
can clearly see in the wavelet coefficients map of Fig.~\ref{plotatrous} 
individual patterns of the Kaiser-Stebbins effect. This shows that
further wavelet 
algorithms can be investigated in order not only to detect the
non-Gaussian signature of cosmic strings globally but also to extract
individual non-Gaussian structures from the map.

\section*{Acknowledgements}
RBB acknowledges financial support from the PPARC in the form of a
research grant. RBB also thanks Klaus Maisinger and Pia Mukherjee for
useful discussions regarding wavelet algorithms.

\end{document}